\documentclass[12pt,a4paper,final]{iopart}

\usepackage{iopams}  
\usepackage{graphicx}
\usepackage[breaklinks=true,colorlinks=true,linkcolor=blue,urlcolor=blue,citecolor=blue]{hyperref}

\begin{document}

\title{Phantom wormholes in Einstein-Maxwell-dilaton theory}

\author{Prieslei Goulart$^{1}$}
\address{$^1$Instituto de F\'{i}sica Te\'{o}rica, UNESP-Universidade Estadual Paulista\\ R. Dr. Bento T. Ferraz 271, Bl. II, S\~{a}o Paulo 01140-070, SP, Brazil}
\ead{prieslei@ift.unesp.br}

\begin{abstract}
In this paper we give an electrically charged traversable wormhole solution for the Einstein-Maxwell-dilaton theory when the dilaton is a phantom field, i.e. it has flipped sign kinetic term appearing in the action. In the limit when the charge is zero, we recover the anti-Fisher solution, which can be reduced to the Bronnikov-Ellis solution under certain choices of integration constants. The equations of motion of this theory share the same S-duality invariance of string theory, so the electrically charged solution is rotated into the magnetically charged one by applying such transformations. The scalar field is topological, so we compute its topological charge, and discuss that under appropriate boundary conditions we can have a lump, a kink, or an anti-kink profile. We determine the position of the throat, and show the embedding diagram of the wormhole. As a physical application, we apply the Gauss-Bonnet theorem to compute the deflection angle of a light-ray that passes close to the wormhole.

\end{abstract}

\vspace{2pc}
\noindent{\it Keywords}: Wormholes, Exact Solutions, Gauss-Bonnet Theorem

\section{Introduction}
Phantom fields are defined as the fields whose kinetic term appears in the action with flipped sign\footnote{The metric signature used in this paper is $(-,+,+,+)$}, resulting in a negative kinetic energy. The motivation for considering such fields comes from cosmological observations \cite{Hannestad:2005fg,Dunkley:2008ie}, which suggest the existence of a fluid with negative pressure that could be a phantom field, and also from a theoretical point of view, since for a nontrivial kinetic term $P(-\frac{1}{2}(\partial \phi)^{2})$ they give a ghost condensate in a consistent infrared modification of gravity \cite{ArkaniHamed:2003uy}. Although negative energies are in general a sign of instabilities, it is argued that such instabilities can be cured \cite{Piazza:2004df}. In string theory, phantom fields appear in the study of "negative branes" \cite{Vafa:2001qf,Okuda:2006fb} (called also  "topological anti-branes" or "ghost branes"). In the same way as ordinary branes, negative branes are extended objects which give rise to a gauge group $SU(M)$, for a stack of $M$ negative branes on top of each other, where $M$ is a negative Chan-Pathon factor associated to the endpoint of the string. They cancel the effects of ordinary branes. Then, $SU(N|M)$ symmetry can be realized for a stack of $N$ ordinary D-branes, and $M$ negative D-branes. It was argued in \cite{Vafa:2014iua} that, if $\mathcal{N}=4$ $SU(N|M)$ gauge theories exist, they must be holographically dual to AdS$_{5}\times S^{5}$ because they are indistinguishable from $SU(N-M)$ theory to all order in $1/(N-M)$ (for $N>M$). Also, the relation among string dualities, the signature of spacetime, and phantom fields, was carefully studied in \cite{Hull:1998ym}. 

An interesting direction is to find and classify the charged solutions in the presence of phantom fields. Consider for instance the Einstein-Maxwell-dilaton (EMD) theory. By flipping the sign of the kinetic term of the gauge field or the dilaton field, Gibbons and Rasheed showed \cite{Gibbons:1996pd} that it is possible to construct massless black holes and wormholes solutions for the new theories. The author of this paper has shown recently \cite{Goulart:2016nkv} that one can set the value of the dilaton at infinity to a specific imaginary value, and construct massless black holes for the EMD theory, whose observables are all real. Moreover, the same massless solutions were used to construct Einstein-Rosen bridges which satisfy the Null Energy Condition (NEC). More general black hole solutions were found for the case when only the dilaton field is of the phantom type \cite{Clement:2009ai}. For convenience, we follow \cite{Gibbons:1996pd} and call such a theory as Einstein-Maxwell-anti-dilaton (EMaD), due to the fact that the kinetic term of the dilaton has a flipped sign. 

An intriguing aspect about phantom scalar fields is the fact that they give rise to traversable wormhole solutions. A wormhole is a tunnel connecting two different regions of the same spacetime, or two regions of different spacetimes \cite{Bergmann:1957zza,Fisher:1948yn,Bronnikov:1973fh,Ellis:1973yv}. A traversable wormhole is a solution that allows observers to cross it from one region to the other. The surface of minimal area connecting the two regions is called the "throat" of the wormhole. Morris and Thorne have shown \cite{Morris:1988cz} that the matter that keeps the throat open is "exotic", i.e. it does not satisfy the null energy condition (NEC). This means that there is matter with negative energy at the throat of the wormhole, which is classically unacceptable. We know nowadays that violations of the NEC happens in quantum mechanics, which, together with the cosmological observations discussed in the first paragraph, is another motivation for the study of wormholes in the presence of phantom fields. Of course there are other motivations for considering wormholes. One recent investigation, for instance, made use of the cut-and-paste method to construct wormholes in the presence of non-linear Maxwell fields \cite{Hendi:2014uea}. Also in \cite{Clement:2009ai}, a wormhole solution was obtained for the case when the coupling of the phantom-dilaton $\phi$ with the Maxwell field is given by the term $e^{2\lambda \phi}F^{2}$, where $\lambda^{2}>1$. 

In this paper we give an analytical charged traversable wormhole solution for the EMaD theory\footnote{For a numerical study about the role of phantom scalar fields in the gravitational collapse,  see reference \cite{Nakonieczna:2015apa}.} for the case when the coupling of the phantom-dilaton $\phi$ with the Maxwell field is given by the term $e^{-2 \phi}F^{2}$. The solution we present is electrically charged, and, as the equations of motion are invariant under S-duality, we can also obtain the magnetically charged solution by aplying such transformation. In the limit when the electric or magnetic charge is zero we obtain the old known wormhole solution discussed by Bergmann and Leipnik \cite{Bergmann:1957zza} (which is sometimes referred to as anti-Fisher solution \cite{Fisher:1948yn}), which can also be reduced to the Bronnikov-Ellis wormhole solution \cite{Bronnikov:1973fh,Ellis:1973yv}. As a physical application, we use the method introduced by Gibbons and Werner \cite{Gibbons:2008rj} and apply the Gauss-Bonnet theorem to compute the deflection angle of the light passing close to the wormhole.

We present our results as follows. In section 2 we present the EMaD theory, and write the equations of motion. In section 3 we give the electrically charged traversable wormhole solution, and show that it reduces to known solutions in the limit when the charge is zero. We briefly discuss that the magnetically charged solution can be obtained via S-duality rotation. In section 4 we derive an equation that gives the position of the throat of the wormhole. In section 5 we compute the topological charge of the anti-dilaton field, and then, we make the plots for the anti-dilaton, exponential coupling and electric field for the cases when the anti-dilaton is a lump and a kink. In section 6 we just present the embedding diagram for the wormhole. In section 7 we compute the deflection angle of a light ray passing close to the wormhole. In section 8 we conclude.

\section{Einstein-Maxwell-anti-dilaton theory}
The field content of the theory we consider is the metric $g_{\mu\nu}$, a gauge field $A_{\mu}$, and a phantom scalar $\phi$, i.e. a scalar field whose kinetic term in the action has a flipped sign. Such theory is called Einstein-Maxwell-anti-dilaton (EMaD) theory, since this is just the Einstein-Maxwell-dilaton (EMD) theory with a positive kinetic term for the dilaton. The action is written as
\begin{equation} S=\int d^{4}x\sqrt{-g}\left(R+2\partial_{\mu}\phi\partial^{\mu}\phi-e^{-2\phi}F_{\mu\nu}F^{\mu\nu}\right). \label{emad} \end{equation}
We take units in which $(16\pi G_N)\equiv 1$. The field strength has the usual form 
\begin{equation}
F_{\mu\nu}=\partial_{\mu}A_{\nu}-\partial_{\nu}A_{\mu}. 
\end{equation}
The equations of motion for the metric, anti-dilaton,  gauge field, and the Bianchi identities are respectively:
\begin{equation}
R_{\mu\nu}=-2\partial_{\mu}\phi\partial_{\nu}\phi-\frac{1}{2}g_{\mu\nu}e^{-2\phi}F_{\rho\sigma}F^{\rho\sigma}+2e^{-2\phi}F_{\mu\rho}
{F_{\nu}}^{\rho}, \label{riccieq}
\end{equation}
\begin{equation}
\nabla_{\mu}(\partial^{\mu}\phi)-\frac{1}{2}e^{-2\phi}F_{\mu\nu}F^{\mu\nu}=0, \label{adileom}\end{equation}
\begin{equation} \nabla_{\mu}\left(e^{-2\phi}F^{\mu\nu}\right)=0,\label{gaugeeom} 
\end{equation}
\begin{equation}
\nabla_{\left[\mu\right.}F_{\left.\rho\sigma\right]}=0.\label{bianchi}
\end{equation}

\section{Electrically charged wormholes}
We want to study electrically charged wormholes in a static spacetime with spherical symmetry, so we take the following ansatz for the metric
\begin{equation} ds^{2}=-e^{-\lambda}dt^{2}+e^{\lambda}dr^{2}+C^{2}(r)(d\theta^{2}+\sin^{2}\theta d\phi^{2}).\label{metric}  \end{equation}
The resulting equations of motion can be combined and written as
\begin{equation}
\frac{C''}{C}=(\phi')^{2}, \label{segderc} \end{equation}
\begin{equation} (e^{-\lambda}C^{2})''=2,\label{segderl}\end{equation}
\begin{equation} \frac{d}{dr}\left(C^{2}\frac{d}{dr}(e^{-\lambda})\right)=-2C^{2}e^{-2\phi}F_{rt}F^{rt},   \label{simp1s} \end{equation}
\begin{equation} \frac{d}{dr}\left(e^{-\lambda}C^{2}\phi'\right)=C^{2}e^{-2\phi}F_{rt}F^{rt}. \label{simp2s}
\end{equation}
The solution to the gauge field equation (\ref{gaugeeom}) is
\begin{equation} F_{rt}=\frac{Q}{e^{-2\phi}C^{2}}. \end{equation}
Replacing this in the previous equations, we find an analytical wormhole solution given by
\begin{equation}
e^{-\lambda}=\exp\left[\frac{Q^{2}}{2c_{1}^{2}}e^{c_{2}+\frac{2c_{1}}{l}\arctan\left(\frac{r}{l}\right)}-\frac{2(b_{1}-c_{1})}{l}\arctan\left(\frac{r}{l}\right)-(2b_{2}-c_{2})\right],\label{temp} 
\end{equation}
\begin{equation}C^{2}=(r^{2}+l^{2})\exp\left[-\frac{Q^{2}}{2c_{1}^{2}}e^{c_{2}+\frac{2c_{1}}{l}\arctan\left(\frac{r}{l}\right)}+\frac{2(b_{1}-c_{1})}{l}\arctan\left(\frac{r}{l}\right)+(2b_{2}-c_{2})\right], \label{ang}\end{equation}
\begin{equation} \phi  =-\frac{Q^{2}}{4c_{1}^{2}}e^{c_{2}+\frac{2c_{1}}{l}\arctan\left(\frac{r}{l}\right)}+\frac{b_{1}}{l}\arctan\left(\frac{r}{l}\right)+b_{2},\label{dilaton} \end{equation}
\begin{equation} F_{rt}=\frac{Q}{(r^{2}+l^{2})}e^{c_{2}+\frac{2c_{1}}{l}\arctan\left(\frac{r}{l}\right)},\label{elfield}
\end{equation}
with the condition
\begin{equation} c_{1}=b_{1}\pm\sqrt{b_{1}^{2}-l^{2}}. \label{condition} \end{equation}
In this solution, $b_{1}$, $b_{2}$, $c_{1}$ and $c_{2}$ are integration constants. Notice that $b_{1}$ and $c_{1}$ have dimension of length, and $b_{2}$ and $c_{2}$ are dimensionless. The constant $b_{2}$ has no geometrical meaning and it just contributes to the value of the dilaton at the asymptotic regions. The constant $c_{2}$ shapes the profile of the phantom scalar, as will be explained in section 5. This is a real solution valid in the whole spacetime. Apart from the term $(r^{2}+l^{2})$ in (\ref{ang}) and (\ref{elfield}), the solution depends only on exponentials of the inverse of tangent function of the radial coordinate. So, it contains no singularity, as a traversable wormhole must be. The wormhole connects one Minkowski spacetime located at $r=+ \infty $ with another one at $r=- \infty $. For $Q=0$, we recover the anti-Fisher solution \cite{Bergmann:1957zza,Fisher:1948yn}
\begin{equation}
ds^{2}=-e^{-\lambda}dt^{2}+e^{\lambda}dr^{2}+C^{2}(r)(d\theta^{2}+\sin^{2}\theta d\phi^{2})\nonumber \end{equation}
\begin{equation} e^{-\lambda}=\exp\left[-\frac{2(b_{1}-c_{1})}{l}\arctan\left(\frac{r}{l}\right)-(2b_{2}-c_{2})\right], \end{equation}
\begin{equation} C^{2}(r)=(r^{2}+l^{2})\exp\left[\frac{2(b_{1}-c_{1})}{l}\arctan\left(\frac{r}{l}\right)+(2b_{2}-c_{2})\right], \end{equation}
\begin{equation} \phi(r)  = \frac{b_{1}}{l}\arctan\left(\frac{r}{l}\right)+b_{2},
\end{equation}
also with the condition (\ref{condition}). We recover the BE wormhole \cite{Bronnikov:1973fh,Ellis:1973yv} if we set $b_{1}=l$ (which implies $b_{1}=c_{1}$), and also $2b_{2}=c_{2}$. For other solutions involving different interacting theories with Lagrangian $\mathcal{L}\sim R-2(\nabla\phi)^{2}-Z(\phi)F^{2}$, where $Z(\phi)$ are different functions of the scalar field, see references \cite{Bronnikov:1977is, Bronnikov:1979ex}.

For reasons that will be explained below, it is more convenient to express our solution (\ref{temp}), (\ref{ang}), (\ref{dilaton}) and (\ref{elfield}) in terms of the constants $Q$, $l$, $b_{1}$, $b_{2}$, $c_{1}$ and $c_{2}$. In general, static black hole and wormhole solutions are expressed in terms of the asymptotic charges such as the mass $M$, the electric charge $q$, and the dilaton $\Sigma$. For a traversable wormhole we have two asymptotic regions, so we can compute the charges for each asymptotic region and write them in terms of these constants. Using equation (\ref{metap}) derived in appendix A, the wormhole metric can be expressed in the positive asymptotic region, i.e. $r\rightarrow +\infty$, as
\begin{equation} 
ds^{2}\approx -e^{-m_{1}}\left(1-\frac{m_{2}}{r}\right)dt^{2}+e^{m_{1}}\left(1-\frac{m_{2}}{r}\right)^{-1}[dr^{2}+r^{2}(d\theta^{2}+\sin^{2}\theta d\phi^{2})],
\end{equation}
where $m_{1}$ and $m_{2}$ are given by equation (\ref{defm1}) also in appendix A. Making the following scale redefinitions
\begin{equation} t\rightarrow e^{m_{1}/2}\tau, \,\,\, r\rightarrow e^{-m_{1}/2}u,
\end{equation}
the metric becomes
\begin{equation} ds^{2}\approx -\left(1-\frac{m_{2}e^{m_{1}/2}}{u}\right)d\tau^{2}+\left(1-\frac{m_{2}e^{m_{1}/2}}{u}\right)^{-1}[du^{2}+u^{2}(d\theta^{2}+\sin^{2}\theta d\phi^{2})]. 
\end{equation}
The term multiplying the spatial part of the metric can be expanded, and we finally obtain
\begin{equation} 
ds^{2}\approx -\left(1-\frac{m_{2}e^{m_{1}/2}}{u}\right)d\tau^{2}+\left(1+\frac{m_{2}e^{m_{1}/2}}{u}\right)[du^{2}+u^{2}(d\theta^{2}+\sin^{2}\theta d\phi^{2})]. \label{wfwh}
\end{equation}
In the weak-field limit the static metric is expressed as
\begin{equation} 
ds^{2}=-(1+2U_{N})d\tau^{2}+(1-2U_{N})[du^{2}+u^{2}(d\theta^{2}+\sin^{2}\theta d\phi^{2})], \label{wflimit}
\end{equation}
where $U_{N}=-\frac{M}{r}$ is the Newtonian potential, and $M$ is the mass parameter of the gravitational object. Comparing equations (\ref{wfwh}) and (\ref{wflimit}) we identify the mass parameter in the positive asymptotic region as
\begin{equation} M_{+}=\frac{m_{2}e^{m_{1}/2}}{2}. 
\end{equation}
Writing this explicitly we obtain
\begin{equation} 
M_{+}=\left(-b_{1}+c_{1}+\frac{Q^2e^{c_{2}+\frac{c_{1}\pi}{l}}}{2c_{1}}\right)\exp\left[-\frac{Q^2e^{c_{2}+\frac{c_{1}\pi}{l}}}{4c_{1}^{2}}+\frac{(b_{1}-c_{1})\pi}{2l}+\frac{(2b_{2}-c_{2})}{2}\right]. 
\end{equation}
This approximation is valid only for the positive asymptotic region, i.e. $r\rightarrow +\infty$. The result for the negative asymptotic region, i.e. $r\rightarrow -\infty$, is obtained by flipping the signs of the terms containing factors of $\pi$, since $\lim_{r\rightarrow -\infty} \arctan(r/l)=-\pi/2$. So, the mass parameter for the negative asymptotic region is written as
\begin{equation} 
M_{-}=\left(-b_{1}+c_{1}+\frac{Q^2e^{c_{2}-\frac{c_{1}\pi}{l}}}{2c_{1}}\right)\exp\left[-\frac{Q^2e^{c_{2}-\frac{c_{1}\pi}{l}}}{4c_{1}^{2}}-\frac{(b_{1}-c_{1})\pi}{2l}+\frac{(2b_{2}-c_{2})}{2}\right]. 
\end{equation}
The electric charge in the positive asymptotic region $q_{+}$ and in the negative asymptotic region $q_{-}$ are defined through the integral
\begin{equation} 
q_{\pm}=\frac{1}{4\pi}\int_{r\rightarrow \pm \infty}F_{\mu\nu}n^{\mu}m^{\nu}\sqrt{g_{\theta\theta}g_{\phi\phi}}d\theta d\phi,  
\end{equation}
where $m^{\mu}=(1,0,0,0)$ and $n^{\mu}=(0,1,0,0)$. This gives 
\begin{eqnarray}
q_{\pm}&=\frac{1}{4\pi}\int_{r\rightarrow \pm \infty}d\theta d\phi\frac{Q}{(r^{2}+l^{2})}e^{c_{2}+\frac{2c_{1}}{l}\arctan\left(\frac{r}{l}\right)}(r^{2}+l^{2})e^{\lambda}\nonumber \\
&=\left. Qe^{c_{2}+\frac{2c_{1}}{l}\arctan\left(\frac{r}{l}\right)}e^{\lambda}\right|_{r\rightarrow \pm \infty}. 
\end{eqnarray}
This implies that
\begin{equation} q_{\pm}=Q\exp\left[-\frac{Q^{2}}{2c_{1}^{2}}e^{c_{2}\pm\frac{c_{1}\pi}{l}}\pm\frac{b_{1}\pi}{l}+2b_{2}\right].
\end{equation}
The dilaton charges at each region, $\Sigma_{\pm}$, are defined through 
\begin{equation} \phi \approx \phi_{\pm}-\frac{\Sigma_{\pm}}{r}+... \,,
\end{equation}
where $\phi_{\pm}$ is the value of the dilaton at the positive region for plus sign and negative asymptotic region for minus sign. Using the approximations given in appendix A, this gives
\begin{equation} \phi \approx -\frac{Q^2e^{c_{2}\pm\frac{c_{1}\pi}{l}}}{4c_{1}^{2}}\pm\frac{b_{1}\pi}{2l}+b_{2}-\frac{1}{r}\left[\frac{Q^2e^{c_{2}\pm\frac{c_{1}\pi}{l}}}{2c_{1}}+b_{1}\right].
\end{equation}
Then we can extract 
\begin{eqnarray}
\phi_{+}&=-\frac{Q^{2}}{4c_{1}^{2}}e^{c_{2}+\frac{c_{1}\pi}{l}}+b_{2}+\frac{b_{1}\pi}{2l},\label{pdi} \\
\phi_{-}&=-\frac{Q^{2}}{4c_{1}^{2}}e^{c_{2}-\frac{c_{1}\pi}{l}}+b_{2}-\frac{b_{1}\pi}{2l},\label{ndi}\\
\Sigma_{+}&=\frac{Q^2e^{c_{2}+\frac{c_{1}\pi}{l}}}{2c_{1}}+b_{1} ,\label{pdc}\\
\Sigma_{-}&=\frac{Q^2e^{c_{2}-\frac{c_{1}\pi}{l}}}{2c_{1}}+b_{1} .\label{ndc}
\end{eqnarray}
This allows us to rewrite the mass parameters $M_{\pm}$ and the charges $q_{\pm}$ as
\begin{eqnarray}
M_{+}&=(\Sigma_{+}-2b_{1}+c_{1})e^{\phi_{+}-\frac{c_{1}\pi}{2l}-\frac{c_{2}}{2}}, \label{posmass}\\
M_{-}&=(\Sigma_{-}-2b_{1}+c_{1})e^{\phi_{-}+\frac{c_{1}\pi}{2l}+\frac{c_{2}}{2}},\label{negmass}\\
q_{+}&=Qe^{2\phi_{+}},\label{posq}\\
q_{-}&=Qe^{2\phi_{-}}\label{negq}.
\end{eqnarray}
Notice that we can not express the constants $Q$, $b_{1}$, $b_{2}$ and $c_{2}$ solely in terms of the charges at each asymptotic region\footnote{Remind that $c_{1}$ depends on $b_{1}$ and $l$ by (\ref{condition})}. This is the reason why it is more convenient to express all our results in terms of these integration constants instead of the asymptotic charges. We will use these results in the end section 7.

For the analysis that follows, we will need the Ricci tensors and the Ricci scalar for our wormhole solution (\ref{temp}), (\ref{ang}), (\ref{dilaton}) and (\ref{elfield}). Using equations (\ref{temp}) and (\ref{ang}), we compute the Ricci tensors, which are given by
\begin{eqnarray}
R_{tt}=\frac{Q^2}{\left(r^2+l^2\right)^2}  \exp \left[\frac{Q^2}{c_{1}^2} e^{\frac{2 c_{1}
  }{l} \arctan\left(\frac{r}{l}\right)+c_{2}}-\frac{2 (2 b_{1}-3 c_{1})}{l}\arctan
   \left(\frac{r}{l}\right)\right. \nonumber \\ \left.  -4 b_{2}+3 c_{2}\right],
\end{eqnarray}
\begin{eqnarray}
R_{rr}=\frac{-1}{2 c_{1}^2
   \left(l^2+r^2\right)^2}\left[4 c_{1}^2 \left((b_{1}-c_{1})^2+l^2\right)+Q^4 e^{\frac{4 c_{1}}{l}\arctan
   \left(\frac{r}{l}\right)+2 c_{2}}\right.\nonumber \\
   \left. +2 c_{1} Q^2
   (c_{1}-2 b_{1}) e^{\frac{2 c_{1} }{l}\arctan
   \left(\frac{r}{l}\right)+c_{2}}\right],
\end{eqnarray}
\begin{equation}
R_{\theta\theta}=\frac{Q^2}{\left(r^2+l^2\right)} e^{\frac{2 c_{1}}{l} \arctan\left(\frac{r}{l}\right)+c_{2}},
\end{equation}
\begin{equation}
R_{\phi\phi}=R_{\theta\theta}\sin^{2}\theta.
\end{equation}
From these results we can easily obtain the Ricci scalar, an it is written as
\begin{eqnarray}
R=\frac{-1}{2 c_{1}^2 \left(l^2+r^2\right)^2}\left(Q^2 e^{\frac{2 c_{1} }{l}\arctan
   \left(\frac{r}{l}\right)+c_{2}}-2b_{1}c_{1}\right)^{2}
   \nonumber \\ 
   \times \exp
   \left(\frac{2 (c_{1}-b_{1}) }{l}\arctan\left(\frac{r}{l}\right)-2
   b_{2}+\frac{Q^2}{2
   c_{1}^2} e^{\frac{2 c_{1} }{l}\arctan
   \left(\frac{r}{l}\right)+c_{2}}+c_{2}\right). \label{curvature}
\end{eqnarray}
Notice that the Ricci scalar (\ref{curvature}) is finite everywhere in the spacetime, i.e. it does not contain any singularity in the range $-\infty<r<+\infty$. This means that the solution is neither a black hole nor a naked singularity. One can check that other scalar invariants constructed out from the Riemann tensors are also finite everywhere. Also, due to the smoothness of the spacetime, all geodesics are complete. In order to check the energy conditions, we choose orthonormal basis vectors \cite{Morris:1988cz}:
\begin{equation}
{\bf{e}}_{\hat{t}}=e^{\lambda/2}\frac{\partial}{\partial t},\\
{\bf{e}}_{\hat{r}}=e^{-\lambda/2}\frac{\partial}{\partial r},\\
{\bf{e}}_{\hat{\theta}}=\frac{1}{C}\frac{\partial}{\partial \theta},\\
{\bf{e}}_{\hat{\phi}}=\frac{1}{C\sin\theta}\frac{\partial}{\partial \phi}.
\end{equation}
In the hatted coordinated system, the components of the energy momentum tensor are  $T_{\hat{t}\hat{t}}=\rho(r)$, $T_{\hat{r}\hat{r}}=-\tau(r)$, $T_{\hat{\theta}\hat{\theta}}=T_{\hat{\phi}\hat{\phi}}
=p(r)$, where $\rho(r)$ is the energy density measured by the static observer, $\tau(r)$ is the tension per unit area measured in the radial direction, and $p(r)$ is the pressure that is measured in the directions orthogonal to the radial direction. The null energy condition is written as
\begin{equation} T_{\hat{\mu}\hat{\nu}}k^{\hat{\mu}}k^{\hat{\nu}}\geq 0, \label{nullec}
\end{equation}
where the null vector is given by $k^{\hat{\mu}}=(1,1,0,0)$. As we are using units in which $(16 \pi G_{N})=1$, we can use Einstein's equations and the fact that $k_{\hat{\mu}}k^{\hat{\mu}}=0$ in order to show that $T_{\hat{\mu}\hat{\nu}}k^{\hat{\mu}}k^{\hat{\nu}}=2R_{\hat{\mu}\hat{\nu}}k^{\hat{\mu}}k^{\hat{\nu}}$. The term $R_{\hat{\mu}\hat{\nu}}k^{\hat{\mu}}k^{\hat{\nu}}$ is just twice the curvature scalar given by (\ref{curvature}). So, for the wormhole solution (\ref{temp}), (\ref{ang}), (\ref{dilaton}) and (\ref{elfield}), the null energy condition (\ref{nullec}) is not satisfied, since (\ref{curvature}) is a stricly negative function of the radial coordinate. It is important to emphasize that this fact does not depend on any choice of integration constants. The curvature is negative everywhere in the spacetime, and it is finite at the throat. 

Although the dilaton field is of the phantom type, the equations of motion for the theory (\ref{emad}) have the same S-duality invariance of the EMD theory, i.e. the equations of motion are invariant under 
\begin{equation} \phi \rightarrow -\phi, \,\,\, F^{\mu\nu}\rightarrow \frac{\tilde{\epsilon}^{\mu\nu\rho\sigma}}{2\sqrt{-g}}F_{\rho\sigma}. \label{sduality}\end{equation}
Here, $\tilde{\epsilon}^{\mu\nu\rho\sigma}$ is the antisymmetric Levi-Civita symbol, and $\tilde{\epsilon}^{tr\theta\phi}=1$. We can transform our electrically charged solution to a magnetically charged solution simply by applying the S-duality transformation (\ref{sduality}). It would be very interesting though to obtain a traversable wormhole solution for the EMaD theory with both electric and magnetic charges, i.e. a dyonic solution. It is known that the dyonic black hole for the EMD theory (with coupling to the gauge field being $e^{-2\phi}$) contains two horizons \cite{Gibbons:1987ps}, unlike its electrically (or magnetically) charged solution, which contains only one \cite{Garfinkle:1990qj}. In the absence of such dyonic wormhole solution, we can not see whether the spacetime structure changes.    

\section{The throat of the wormhole}
As was stated in the introduction, the throat of the wormhole corresponds to the surface of minimal area. Notice that the wormhole metric can be cast in the form
\begin{equation} ds^{2}=-e^{-\lambda}dt^{2}+e^{\lambda}\left[dr^{2}+(r^{2}+l^{2})d\Omega_{2}^{2}\right].    \end{equation}
The spatial part of the metric would be the same as the BE wormhole case if the factor $e^{-\lambda}$ were absent. In the BE wormhole the minimal surface happens when the radial coordinate is $r=0$, but we will see that this is not necessarily the case here. The function $C^{2}(r)=e^{\lambda}(r^{2}+l^{2})$ has a minimum when 
\begin{equation} (C^{2})'(r_{\min})=0, \,\,\, (C^{2})''(r_{\min})>0. \end{equation}
These two conditions imply that, at the minimum radius 
\begin{equation} 2CC'=0, \,\,\, 2(C')^{2}+2CC''> 0. \label{condmin}\end{equation}
The function $C$ is non-zero everywhere in the wormhole spacetime, so the position of the minimum, $r_{\min}$, can be found solving
\begin{equation} C'(r_{\min})=0, \end{equation}
whereas the second condition implies that the signs of $C(r_{\min})$ and $C''(r_{\min})$ be the same. If $2(C')^{2}+2CC''= 0$, then we must analyse the behavior of the first derivative around the inflection point to check that it is indeed a minimum. Using the first condition we see that the position of the throat $r_{\min}$ is found solving the equation 
\begin{equation} r_{\min}-\frac{Q^{2}}{2c_{1}}e^{c_{2}+\frac{2c_{1}}{l}\arctan\left(\frac{r_{\min}}{l}\right)}+b_{1}-c_{1}=0. \label{minradius}\end{equation}
Notice that $r_{\min}$ can have different values because the inverse tangent is a multi-valued function. We focus our analysis considering only the principal value of the inverse tangent, so the argument inside the tangent function ranges from
\begin{equation} \frac{r}{l}=\tan y\rightarrow -\frac{\pi}{2}<y<\frac{\pi}{2}. \label{branch}\end{equation}
Notice that our solution depends on several integration constants, but, as we are interested only in principal values, equation (\ref{minradius}) gives one single value for $r_{\min}$. One can easily plot the function $C^{2}(r)$ and check that this value is indeed a minimum.

\section{Topological charge and plots}
\begin{figure}[h]
\centering 
\includegraphics[width=8cm, height=6cm]{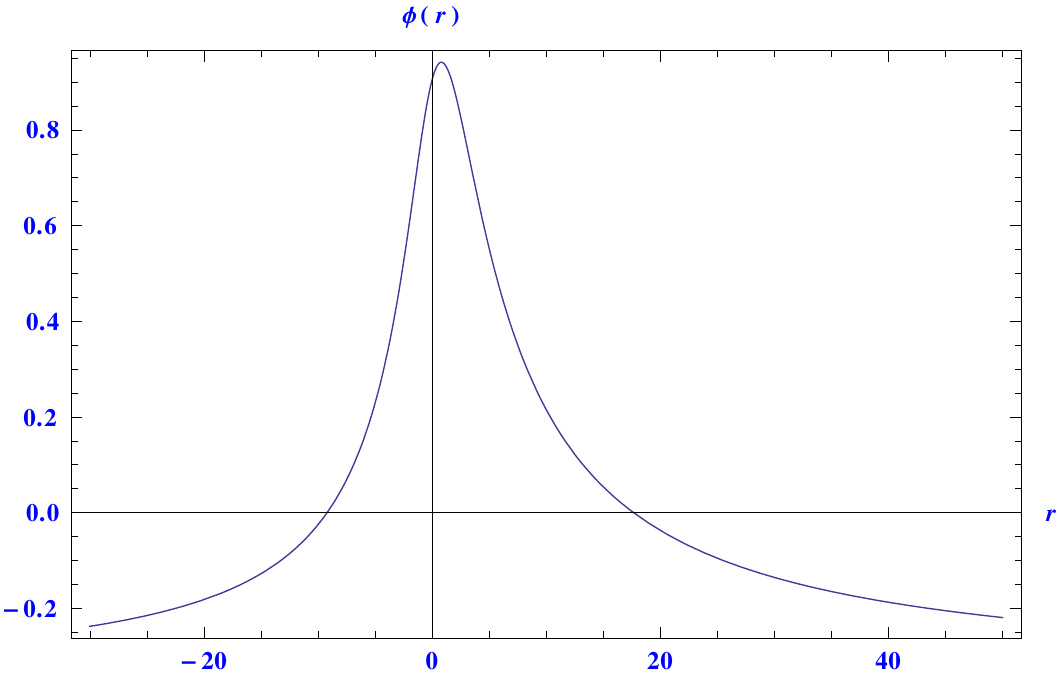}\\
\caption{Dilaton field. $Q=2, \, b_{1}=5, \, l=3, \, c_{1}=1,\,b_{2}=3$.}\label{fig1}
\end{figure}

\begin{figure}[h]
\centering 
\includegraphics[width=8cm, height=6cm]{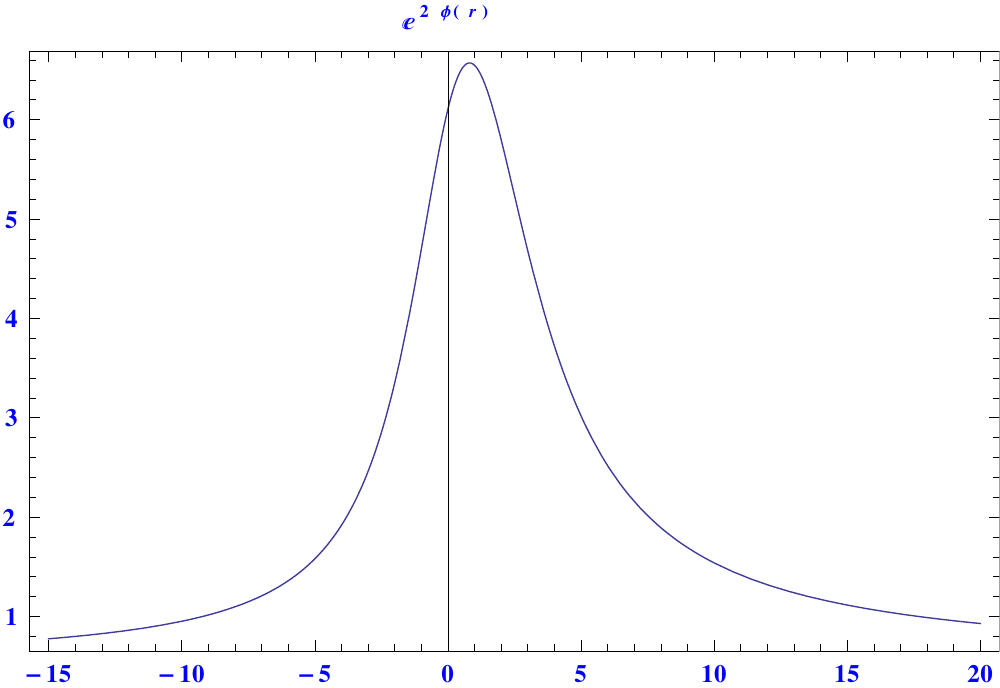}\\
\caption{Exponential coupling $e^{2\phi}$. $Q=2, \, b_{1}=5, \, l=3, \, c_{1}=1,\,b_{2}=3$.}\label{fig2}
\end{figure}

\begin{figure}[h]
\centering 
\includegraphics[width=8cm, height=6cm]{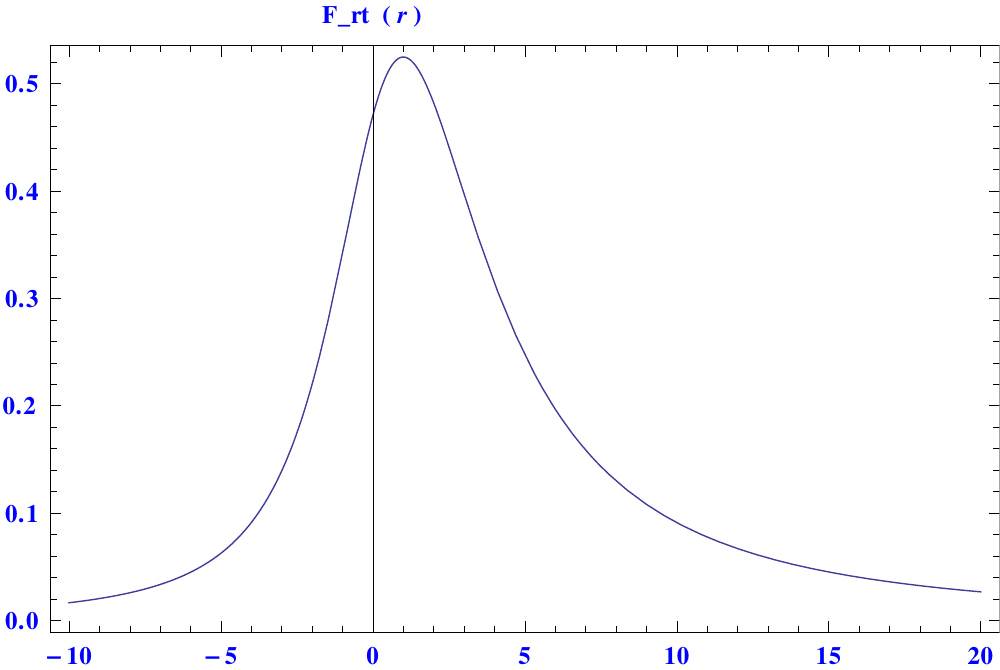}\\
\caption{Electric field. $Q=2, \, b_{1}=5, \, l=3, \, c_{1}=1,\,b_{2}=3$.}\label{fig3}
\end{figure}

\begin{figure}[h]
\centering 
\includegraphics[width=8cm, height=6cm]{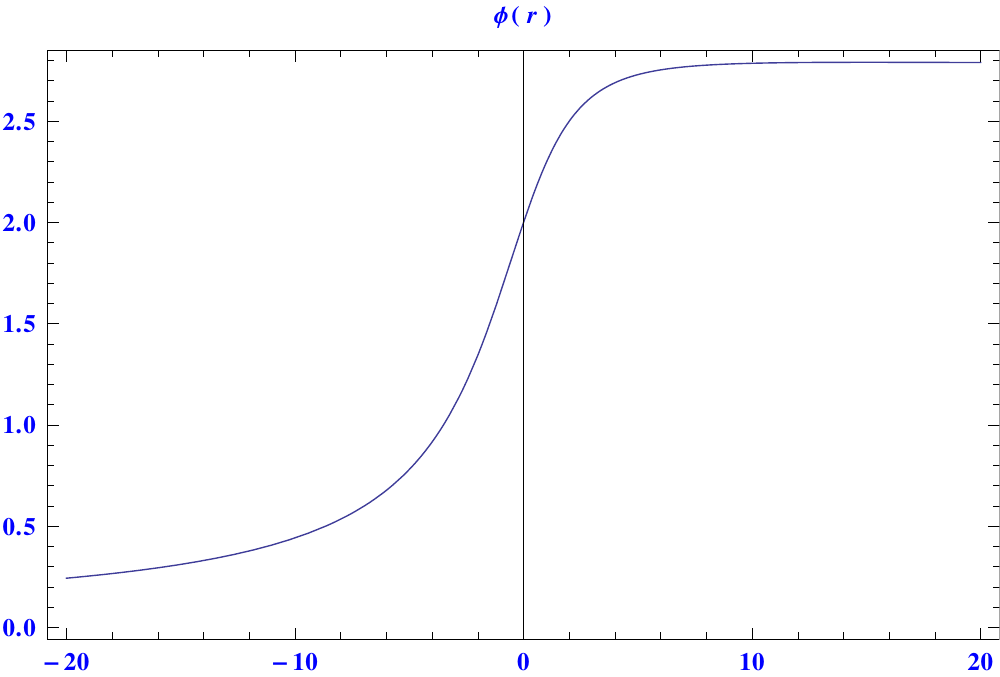}\\
\caption{Dilaton field. $c_{2}=0, \, Q=2, \, b_{1}=5, \, l=3, \, c_{1}=1,\,b_{2}=3$.}\label{fig4}
\end{figure}

\begin{figure}[h]
\centering 
\includegraphics[width=8cm, height=6cm]{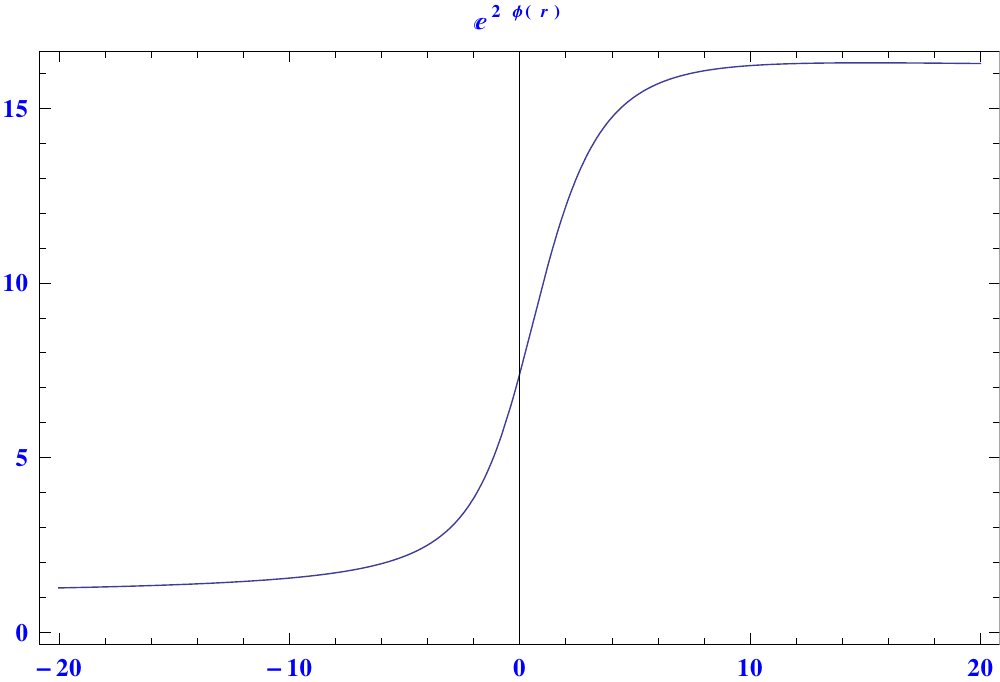}\\
\caption{Exponential coupling $e^{2\phi}$. $c_{2}=0, \, Q=2, \, b_{1}=5, \, l=3, \, c_{1}=1,\,b_{2}=3$.}\label{fig5}
\end{figure}

\begin{figure}[h]
\centering 
\includegraphics[width=8cm, height=6cm]{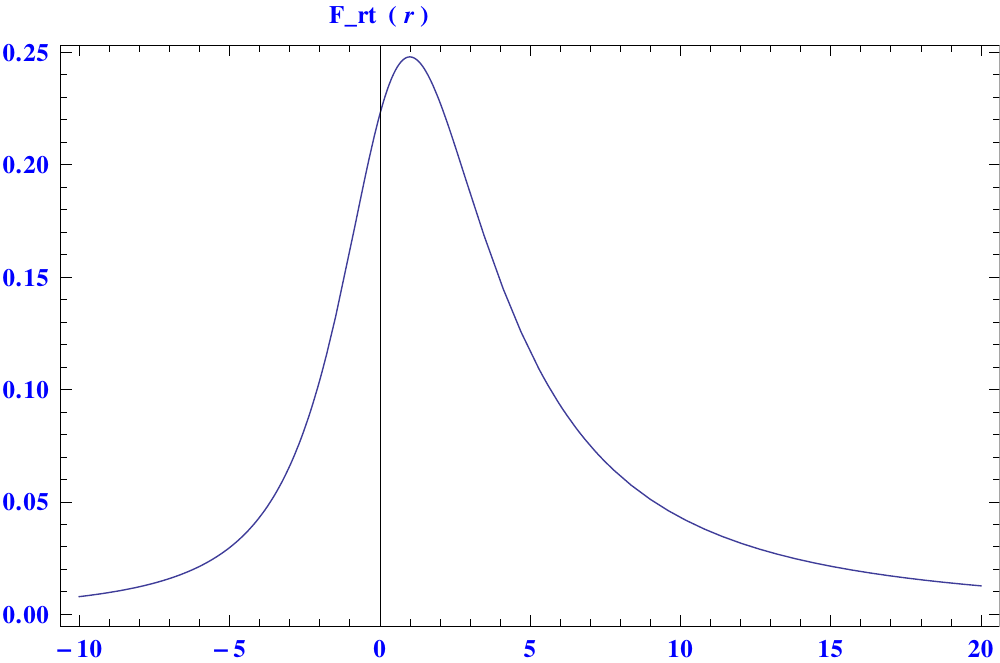}\\
\caption{Electric field. $c_{2}=0, \, Q=2, \, b_{1}=5, \, l=3, \, c_{1}=1,\,b_{2}=3$.}\label{fig6}
\end{figure}

A topological charge is a conserved quantity that is not associated to any Noether symmetry. The field $\phi$ given by equation (\ref{dilaton}) is of topological nature, so it must have a topological charge associated to it. In fact, we can define the following current
\begin{equation} j^{\mu}\sim \tilde{\epsilon}^{\mu\nu}\partial_{\nu}\phi, \end{equation}
such that
\begin{equation} N=\beta \int_{-\infty}^{+\infty} dr\tilde{\epsilon}^{tr}\partial_{r}\phi=\beta[\phi(+\infty)-\phi(-\infty)], \end{equation}
for a constant $\beta$, which we will fix as $\beta=1$, and a field $\phi$ that depends only on the radial coordinate. Here, $\tilde{\epsilon}^{\mu\nu}$ is the antisymmetric Levi-Civita symbol with two indices, and $\tilde{\epsilon}^{tr}=1$. For $l>0$ we have 
\begin{equation} \lim_{r\to\pm\infty} \arctan\left(\frac{r}{l}\right)=\pm\frac{\pi}{2}(1+4n), \,\,\, n=0,1,2,... . \end{equation}
We choose the branch for which $n=0$ in order to be consistent with the choice of interval in equation (\ref{branch}). The dilaton field (\ref{dilaton}) at the two asymptotic regions is written as
\begin{equation} \phi(\pm\infty)= -\frac{Q^{2}}{4c_{1}^{2}}e^{c_{2}\pm\frac{c_{1}\pi}{l}}+b_{2}\pm\frac{b_{1}\pi}{2l}. \end{equation}
So, the topological charge is written as
\begin{equation} N=\phi(+\infty)-\phi(-\infty)= -\frac{Q^{2}}{2c_{1}^{2}}e^{c_{2}}\sinh\left(\frac{c_{1}\pi}{l}\right)+\frac{b_{1}\pi}{l}. \end{equation}
The factor $c_{2}$ is important because it shapes the profile of the scalar field. Depending on the choice of integration constants, the dilaton can be connected to two different vacua, i.e. it is a kink, or it can be connected to the same vacuum, i.e. it is a lump. In order to obtain a lump, the topological charge must be zero, which implies 
\begin{equation} \phi(\infty)=\phi(-\infty)\Rightarrow e^{c_{2}}=\frac{2c_{1}^{2}b_{1}\pi}{Q^{2}l}\frac{1}{\sinh{\left(\frac{c_{1}\pi}{l}\right)}}. \label{condlump}\end{equation}
We first plot the dilaton field for the case when it is a lump, and this is shown in figure \ref{fig1}. For the same values of constants, the coupling $e^{2\phi}$ also has a lump profile, as shown in figure \ref{fig2}. For completeness, the electric field is also plotted in figure \ref{fig3}. For values of $c_{2}$ other than those in (\ref{condlump}), the dilaton will have a kink or anti-kink profile. We plot only the case when $c_{2}=0$. The dilaton is shown in figure \ref{fig4}. Notice now, that the exponential coupling is also a kink, as shown in figure \ref{fig5}. The electric field is also plotted in figure \ref{fig6}.

\section{Embedding diagram}
\begin{figure}[h]
\centering 
\includegraphics[width=8cm, height=7cm]{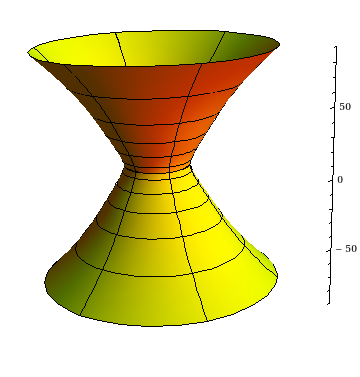}\\
\caption{Embedding diagram: $Q=2, \, b_{1}=5, \, l=3, \, c_{1}=1,\,b_{2}=3, \, e^{c_{2}}=(3\sqrt{3}+4)/(2e^{2\pi/9})$.}\label{ed}
\end{figure}

In this section we construct the embedding diagram for the wormhole. Without loss of generality, we set $t=$const and $\theta=\pi/2$. The metric for this slice is then written as
\begin{equation} ds^{2}=e^{\lambda}(dr^{2}+(r^{2}+l^{2})d\phi^2). \label{whsimp}\end{equation}
This differs from the usual embedding diagram of the BE wormhole only by a "conformal factor" $e^{\lambda}$. We redefine our radial coordinate such that
\begin{equation} (r^{*})^{2}\equiv e^{\lambda}(r^{2}+l^{2}). \end{equation}
We can not invert this equation in order to write $r(r^{\star})$. Taking derivatives, we have
\begin{equation} dr^{*}=\left(r-\frac{Q^{2}}{2c_{1}}e^{c_{2}+\frac{2c_{1}}{l}\arctan\left(\frac{r}{l}\right)}+b_{1}-c_{1}\right)\frac{e^{\lambda/2}}{\sqrt{r^{2}+l^{2}}} dr. \end{equation}
For simplicity, we redefine the term inside the parenthesis as
\begin{equation} g(r^{*})\equiv r-\frac{Q^{2}}{2c_{1}}e^{c_{2}+\frac{2c_{1}}{l}\arctan\left(\frac{r}{l}\right)}+b_{1}-c_{1}, \end{equation}
where it is implicit the relation $r(r^{\star})$. Then, (\ref{whsimp}) is rewritten as
\begin{equation} ds^{2}=\frac{r^{2}+l^{2}}{g^{2}}dr^{*2}+r^{*2}d\phi^2. \label{whsimp1}\end{equation}
The Euclidean metric of the embedding space is the same as the one used in \cite{Morris:1988cz}
\begin{equation} ds^{2}=dz^{2}+dr^{*2}+r^{*2}d\phi^2=\left[1+\left(\frac{dz}{dr*}\right)^{2}\right]dr^{*2}+r^{*2}d\phi^{2}. \end{equation}
This implies that
\begin{equation} \frac{dz}{dr*}=\pm \left( \frac{r^{2}+l^{2}}{g^{2}}-1 \right)^{1/2}. \end{equation}
We split the function $z(r^{\star})$ in two parts, one for the positive sign, $z_{+}(r^{\star})$, and another one for the negative sign, $z_{-}(r^{\star})$. So we have two differential equations written as
\begin{equation} \frac{dz_{+}}{dr*}=+ \left( \frac{r^{2}+l^{2}}{g^{2}}-1 \right)^{1/2}, \,\,\, \frac{dz_{-}}{dr*}=- \left( \frac{r^{2}+l^{2}}{g^{2}}-1 \right)^{1/2}. \end{equation}
These equations are integrated numerically, and the embedding diagram is shown in figure (\ref{ed}). The boundary condition we imposed is just $z_{+}(r_{\min})=z_{-}(r_{\min})=0$. This diagram represents the wormhole, and it connects the upper region $z_{+}(r^{\star})$ with the lower region $z_{-}(r^{\star})$ by a minimal surface with area greater than zero.

\section{Deflection angle via Gauss-Bonnet theorem}
In a geometrical approach to gravitational lensing theory, Gibbons and Werner showed how the Gauss-Bonnet theorem can be applied to the computation of the light deflection angle in the weak deflection limit for static and spherically symmetric spacetimes \cite{Gibbons:2008rj}. The application of this method for wormhole cases was done in references \cite{Jusufi:2017gyu,Jusufi:2017vta}. In this section, we apply this method to compute the deflection angle of a light ray passing close to the wormhole described by the solution of section 3. 

\begin{figure}[h]
\centering 
\includegraphics[width=12cm, height=9cm]{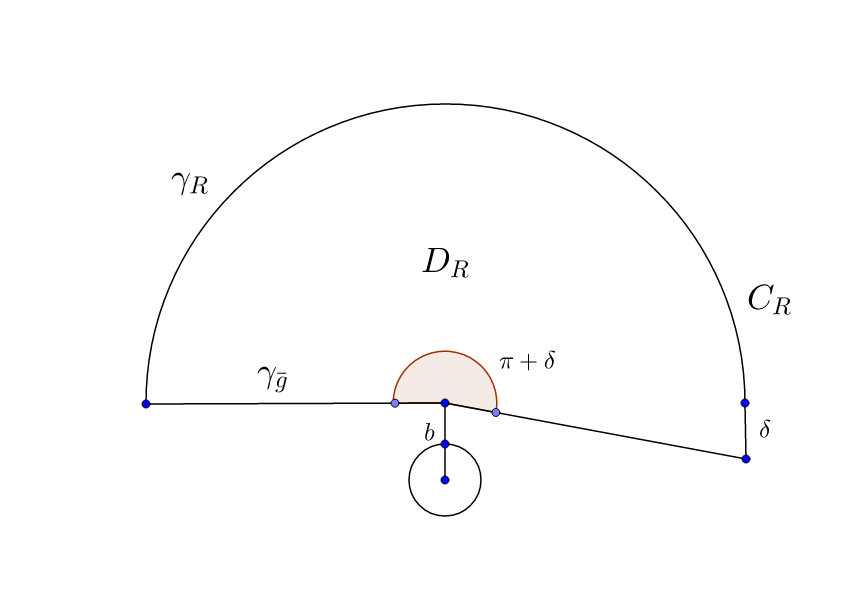}\\
\caption{Gravitational lens: $D_{R}$ is the domain enclosed by the boundary $C_{R}$, $b$ is the impact parameter, and $\delta$ is the deflection angle.}\label{lens}
\end{figure}

Consider the following oriented surface domain $\mathcal{D}$, with boundary $C_{R}=\gamma_{R}\cup \gamma_{\bar{g}}$, as described by figure (\ref{lens}). The Gauss-Bonnet theorem states that
\begin{equation} \int \int_{\mathcal{D}}KdS+\int_{\partial \mathcal{D}} \kappa dt+\sum_{i}\alpha_{i}=2\pi \chi(\mathcal{D}), \label{GB} \end{equation}
where $K$ is the Gaussian curvature associated to the Riemannian metric $\bar{g}$, $\kappa$ is the geodesic curvature for $C_{R}: \{t\}\rightarrow \mathcal{D}$, and $\alpha_{i}$ is the extrerior angle with $i^{th}$ vertex. $\chi(\mathcal{D})$ is the Euler characteristic, which is $\chi(\mathcal{D})=1$ for a non-singular domain, and $\chi(\mathcal{D})=0$ for a singular domain \cite{Gibbons:2008rj}. As the wormhole spacetime is non-singular, we must use $\chi(\mathcal{D})=1$. In order to find the Gaussian curvature, we consider without loss of generality only null geodesics $ds^{2}=0$ on the equatorial plane $\theta=\pi/2$. Then, the wormhole metric reduces to
\begin{equation} dt^{2}=\bar{g}_{ij}dx^{i}dx^{j}=e^{2\lambda}dr^{2}+e^{2\lambda}(r^{2}+l^{2})d\varphi^{2}.\label{optical} \end{equation}
This is the so-called optical metric. We can easily obtain the Christoffel connections for this metric, which are given by
\begin{equation}
\bar{\Gamma}^{r}_{\varphi\varphi}=-r+\frac{Q^{2}}{c_{1}}e^{c_{2}+\frac{2c_{1}}{l}\arctan\left(\frac{r}{l}\right)}-2(b_{1}-c_{1}), \label{con1} \end{equation}
\begin{equation} \bar{\Gamma}^{r}_{r\varphi}=\frac{1}{(r^{2}+l^{2})}\left[r-\frac{Q^{2}}{c_{1}}e^{c_{2}+\frac{2c_{1}}{l}\arctan\left(\frac{r}{l}\right)}+2(b_{1}-c_{1})\right]. 
\end{equation}
We will use these connections later. We introduce the Regge-Wheeler tortoise coordinate $r^{\star}$, such that
\begin{equation} dr^{\star}=e^{\lambda}dr, \,\,\, f^{2}(r^{\star})=e^{2\lambda}(r^{2}+l^{2}),  \label{defcoord}\end{equation}
where $r=r(r^{\star})$. The optical metric (\ref{optical}) then takes the form
\begin{equation} dt^{2}=dr^{\star 2}+f^{2}(r^{\star})d\varphi^{2}. \end{equation}
The Gaussian curvature $K$ is related to the Riemann tensor through the relation
\begin{equation} R_{r\varphi r\varphi}=K(\bar{g}_{r\varphi}\bar{g}_{\varphi r}-\bar{g}_{rr}\bar{g}_{\varphi\varphi})=-K det\bar{g}.  \end{equation}
The Gaussian curvature can be expressed as \cite{Gibbons:2008rj}
\begin{equation} K=-\frac{R_{r\varphi r\varphi}}{det\bar{g}}=-\frac{1}{f(r^{\star})}\frac{d^{2}f(r^{\star})}{dr^{\star 2}}. \end{equation}
Notice that we can write this expression in terms of the original radial coordinate $r$ as 
\begin{equation}
K=-\frac{1}{f(r^{\star})}\left[ \frac{dr}{dr^{\star}}\frac{d}{dr}\left(\frac{dr}{dr^{\star}}\right)\frac{df}{dr}+ \left(\frac{dr}{dr^{\star}}\right)^{2}\frac{d^{2}f}{dr^{2}}\right].
\end{equation}
Using (\ref{defcoord}) we obtain
\begin{equation}
K=-\frac{1}{e^{2\lambda}(r^{2}+l^{2})}\left[\lambda' r +\frac{l^{2}}{(r^{2}+l^{2})}+(r^{2}+l^{2})\lambda''\right]\nonumber \end{equation}
\begin{equation} =-\frac{1}{e^{2\lambda}(r^{2}+l^{2})^{2}}\left[\frac{Q^{2}}{c_{1}}re^{c_{2}+\frac{2c_{1}}{l}\arctan\left(\frac{r}{l}\right)}-2Q^{2}e^{c_{2}+\frac{2c_{1}}{l}\arctan\left(\frac{r}{l}\right)}-2r(b_{1}-c_{1})+l^{2}  \right].
\end{equation}
For the metric (\ref{optical}), we have $\sqrt{\bar{g}}=e^{2\lambda}\sqrt{r^{2}+l^{2}}$. We will compute the first integral in (\ref{GB}) only to leading order for large values of $r$, i.e.
\begin{equation}
\int \int_{D}KdS  =\int_{0}^{\pi}\int_{\frac{b}{\sin\varphi}}^{\infty}dr d\varphi\sqrt{det\bar{g}}K \nonumber \end{equation}
\begin{equation} =\int_{0}^{\pi}\int_{\frac{b}{\sin\varphi}}^{\infty}dr d\varphi e^{2\lambda}\sqrt{r^{2}+l^{2}} K\nonumber \end{equation}
\begin{equation}\approx -  \int_{0}^{\pi}\int_{\frac{b}{\sin\varphi}}^{\infty}dr d\varphi \frac{1}{r^{2}}\left[ \frac{Q^{2}}{c_{1}}e^{c_{2}+\frac{c_{1}\pi}{l}}-2(b_{1}-c_{1})\right]\nonumber \end{equation}
\begin{equation}\approx -\frac{2}{b}\left[ \frac{Q^{2}}{c_{1}}e^{c_{2}+\frac{c_{1}\pi}{l}}-2(b_{1}-c_{1})\right],
\label{int1}\end{equation}
where $b$ is the impact parameter. The second integral is a little more involved. In the examples presented in \cite{Gibbons:2008rj} the result of the integral is always $\pi+\delta$, but this is not always the case. In fact, there are some examples of solutions in the presence of topological defects which give a different contribution \cite{Jusufi:2015laa}, which will be the situation here. In order to compute the second integral, we must find $\kappa$. Let us define the velocity and acceleration vectors along the curve $\gamma$ respectively as $\dot{\gamma}$ and $\ddot{\gamma}$. The velocity vector must respect the unit velocity condition $\bar{g}(\dot{\gamma},\dot{\gamma})=1$. For very large values of $R$ the sum of the external angles for the source and the observer tends to $\pi$, i.e. $\alpha_{S}+\alpha_{O}\rightarrow \pi $. The geodesic curvature is computed using the relation
\begin{equation} \kappa = \bar{g}(\nabla_{\dot{\gamma}}\dot{\gamma},\ddot{\gamma}). \end{equation}
Along $\gamma_{\bar{g}}$, $\kappa(\gamma_{\bar{g}})=0$ because $\gamma_{\bar{g}}$ is a geodesic. We must then compute
\begin{equation} \kappa(\gamma_{R})=|\nabla_{\dot{\gamma}_{R}}\dot{\gamma}_{R}|,  \end{equation}
where $\dot{\gamma}_{R}$ is the velocity vector along the curve $\gamma_{R}$. The radial component of this expression is given by
\begin{equation} (\nabla_{\dot{\gamma}_{R}}\dot{\gamma}_{R})^{r}=\dot{\gamma}^{\mu}_{R}\partial_{\mu}\dot{\gamma}_{R}^{r}+ \dot{\gamma}^{\mu}_{R}\bar{\Gamma}^{r}_{\mu\nu}\dot{\gamma}_{R}^{\nu}.\end{equation}
Because we confine our attention on the countour $C_{R}:=r(\phi)=R=$const for large $R$, there is no variation in the radial distance, so $\dot{\gamma}_{R}^{r}=0$, and $\bar{g}_{\varphi\varphi}(\dot{\gamma}^{\varphi}_{R})^{2}=1$. So, the expression above is written as
\begin{equation} (\nabla_{\dot{\gamma}_{R}}\dot{\gamma}_{R})^{r}= \bar{\Gamma}^{r}_{\varphi\varphi}(\dot{\gamma}_{R}^{\phi})^{2}.\end{equation}
Using (\ref{con1}) we have
\begin{equation} 
(\nabla_{\dot{\gamma}_{R}}\dot{\gamma}_{R})^{r}=\frac{1}{e^{2\lambda}(r^{2}+l^{2})} \left(-r+\frac{Q^{2}}{c_{1}}e^{c_{2}+\frac{2c_{1}}{l}\arctan\left(\frac{r}{l}\right)}-2(b_{1}-c_{1})\right).
\label{compkappa}\end{equation}
Using the optical metric (\ref{optical}), we also obtain
\begin{equation} dt=e^{\lambda}\sqrt{r^{2}+l^{2}}d\varphi. \label{dt} \end{equation}
Combining both results (\ref{compkappa}) and (\ref{dt}), and also using the expansions written explicitly in the appendix, we evaluate the integral as
\begin{equation} \int_{\partial D} \kappa dt= \int_{0}^{\pi+\delta}(\nabla_{\dot{\gamma}_{R}}\dot{\gamma}_{R})^{r}d\varphi\approx -e^{-m_{1}}(\pi+\delta), \label{int2} \end{equation}
where $m_{1}$ is given by (\ref{defm1}). This is valid for large values of $R$ and for small values of $\delta$. Inserting the results (\ref{int1})  and (\ref{int2}) in (\ref{GB}), we can easily evaluate the light deflection angle, given by
\begin{equation} \delta = \frac{2e^{m_{1}}}{b}\left[ 2(b_{1}-c_{1})-\frac{Q^{2}}{c_{1}}e^{c_{2}+\frac{c_{1}\pi}{l}}\right]-\pi(1+e^{m_{1}}).
\end{equation}
This result was derived only for the positive region, but we can easily generalize it for the negative region as well. In fact, using the asymptotic values of the dilaton (\ref{pdi}) and (\ref{ndi}), and the definition of the asymptotic charges  (\ref{pdc}), (\ref{ndc}), (\ref{posmass}), (\ref{negmass}), (\ref{posq}), and (\ref{negq}), the deflection angle in both regions can also be written as
\begin{equation} \delta_{+}=-\frac{4M_{+}}{b}e^{\phi_{+}-\frac{c_{1}\pi}{2l}-\frac{c_{2}}{2}}-\pi(1+e^{2\phi_{+}-\frac{c_{1}\pi}{l}-c_{2}}),
\end{equation}
\begin{equation} \delta_{-}=-\frac{4M_{-}}{b}e^{\phi_{-}+\frac{c_{1}\pi}{2l}+\frac{c_{2}}{2}}-\pi(1+e^{2\phi_{-}+\frac{c_{1}\pi}{l}+c_{2}}). 
\end{equation}

\section{Conclusions}
In this paper we gave an analytical electrically charged traversable wormhole solution for the Eintein-Maxwell-anti-dilaton theory. We discussed that the equations of motion inherit the same invariance under S-duality (electric-magnetic) transformation of string theory, and so, the magnetically charged solution is obtained easily by applying such transformations in the solution. The solution depends on functions of the inverse tangent function of the radial coordinate. This introduces branch cuts which can be absorbed by redefinition of the integration constants, or by considering only principal values  of the inverse tangent. We gave an explicit equation that determines the position of the throat of the wormhole. We computed the topological charge of the anti-dilaton, and discussed that under appropriate boundary conditions it can be a lump, a kink, or anti-kink. Finally, we used the Gauss-Bonnet theorem to compute the deflection angle of a light-ray passing close to this wormhole.

\vspace{10pt}

{\bf{Acknowledgments}}\\
We thank Horatiu Nastase for important comments on the manuscript, and Pedro Vieira for his help with Mathematica. The work of PG is supported by FAPESP grant 2013/00140-7.

\appendix

\section{Details of the approximation}
In this appendix we present explicitly the detail of the approximation used to obtain the result (\ref{int2}). The series expansion for the inverse tangent is written as
\begin{equation} \arctan\left(\frac{x}{l}\right)=\sum_{n=0}^{\infty}\frac{(-1)^{n}}{2n+1}\left(\frac{x}{l}\right)^{2n+1}, \,\,\, \left|\frac{x}{l}\right|<1. \end{equation}
Notice that the limit for when the expansion is valid corresponds to $|x|<l$. We are interested in the limit when $|r|>l$, so we must use the following identity and corresponding expansion
\begin{equation} \arctan\left(\frac{r}{l}\right)= \frac{\pi}{2}-\arctan\left(\frac{l}{r}\right)\approx \frac{\pi}{2} -\frac{l}{r}+\frac{l^{3}}{3r^{3}}-\frac{l^{5}}{5r^{5}}+\frac{l^{7}}{7r^{7}}+... \end{equation}
The exponential of the inverse tangent has then the following expansion
\begin{equation}
\exp\left[\frac{2c_{1}}{l}\arctan\left(\frac{r}{l}\right)\right]\approx \exp \left[\frac{2c_{1}}{l}\left(\frac{\pi}{2}-\frac{l}{r}\right)\right]\nonumber \end{equation}
\begin{equation} \approx e^{\frac{c_{1}\pi}{l}} \left(1-\frac{2c_{1}}{r}\right), \end{equation}
\begin{equation}\frac{2(b_{1}-c_{1})}{l}\arctan\left(\frac{r}{l}\right)\approx \frac{(b_{1}-c_{1})\pi}{l}-\frac{2(b_{1}-c_{1})}{r}.
\end{equation}
These will be enough to expand the function $\lambda$ for large $r$, which results in
\begin{equation} \lambda \approx -\frac{Q^2e^{c_{2}+\frac{c_{1}\pi}{l}}}{2c_{1}^{2}}+\frac{(b_{1}-c_{1})\pi}{l}+2b_{2}-c_{2}+\left(-2b_{1}+2c_{1}+\frac{Q^2e^{c_{2}+\frac{c_{1}\pi}{l}}}{c_{1}}\right)\frac{1}{r}. \end{equation}
Defining
\begin{equation}
m_{1}\equiv -\frac{Q^2e^{c_{2}+\frac{c_{1}\pi}{l}}}{2c_{1}^{2}}+\frac{(b_{1}-c_{1})\pi}{l}+2b_{2}-c_{2}, \;\;\; m_{2}\equiv -2b_{1}+2c_{1}+\frac{Q^2e^{c_{2}+\frac{c_{1}\pi}{l}}}{c_{1}},
\label{defm1}\end{equation}
we have
\begin{equation}
\lambda \approx m_{1}+\frac{m_{2}}{r},
\end{equation}
\begin{equation}e^{-\lambda}\approx e^{-m_{1}}\left(1-\frac{m_{2}}{r}\right).\label{metap}
\end{equation}
In the text, we used only the leading term in this expansion.

\section*{References}



\providecommand{\href}[2]{#2}\begingroup\raggedright\endgroup


\begin{thebibliography}{10}

\bibitem{Hannestad:2005fg}
S.~Hannestad, ``{Dark energy and dark matter from cosmological observations},''
  \href{http://dx.doi.org/10.1142/S0217751X06032885}{{\em Int. J. Mod. Phys.}
  {\bf A21} (2006)  1938--1949},
  \href{http://arxiv.org/abs/astro-ph/0509320}{{\tt arXiv:astro-ph/0509320
  [astro-ph]}}.
[,364(2005)].

\bibitem{Dunkley:2008ie}
{\bf WMAP} Collaboration, J.~Dunkley {\em et al.}, ``{Five-Year Wilkinson
  Microwave Anisotropy Probe (WMAP) Observations: Likelihoods and Parameters
  from the WMAP data},''
  \href{http://dx.doi.org/10.1088/0067-0049/180/2/306}{{\em Astrophys. J.
  Suppl.} {\bf 180} (2009)  306--329},
\href{http://arxiv.org/abs/0803.0586}{{\tt arXiv:0803.0586 [astro-ph]}}.

\bibitem{ArkaniHamed:2003uy}
N.~Arkani-Hamed, H.-C. Cheng, M.~A. Luty, and S.~Mukohyama, ``{Ghost
  condensation and a consistent infrared modification of gravity},''
  \href{http://dx.doi.org/10.1088/1126-6708/2004/05/074}{{\em JHEP} {\bf 05}
  (2004)  074},
\href{http://arxiv.org/abs/hep-th/0312099}{{\tt arXiv:hep-th/0312099
  [hep-th]}}.

\bibitem{Piazza:2004df}
F.~Piazza and S.~Tsujikawa, ``{Dilatonic ghost condensate as dark energy},''
  \href{http://dx.doi.org/10.1088/1475-7516/2004/07/004}{{\em JCAP} {\bf 0407}
  (2004)  004},
\href{http://arxiv.org/abs/hep-th/0405054}{{\tt arXiv:hep-th/0405054
  [hep-th]}}.

\bibitem{Vafa:2001qf}
C.~Vafa, ``{Brane / anti-brane systems and U(N|M) supergroup},''
\href{http://arxiv.org/abs/hep-th/0101218}{{\tt arXiv:hep-th/0101218
  [hep-th]}}.

\bibitem{Okuda:2006fb}
T.~Okuda and T.~Takayanagi, ``{Ghost D-branes},''
  \href{http://dx.doi.org/10.1088/1126-6708/2006/03/062}{{\em JHEP} {\bf 03}
  (2006)  062},
\href{http://arxiv.org/abs/hep-th/0601024}{{\tt arXiv:hep-th/0601024
  [hep-th]}}.

\bibitem{Vafa:2014iua}
C.~Vafa, ``{Non-Unitary Holography},''
\href{http://arxiv.org/abs/1409.1603}{{\tt arXiv:1409.1603 [hep-th]}}.

\bibitem{Hull:1998ym}
C.~M. Hull, ``{Duality and the signature of space-time},''
  \href{http://dx.doi.org/10.1088/1126-6708/1998/11/017}{{\em JHEP} {\bf 11}
  (1998)  017},
\href{http://arxiv.org/abs/hep-th/9807127}{{\tt arXiv:hep-th/9807127
  [hep-th]}}.

\bibitem{Gibbons:1996pd}
G.~W. Gibbons and D.~A. Rasheed, ``{Dyson pairs and zero mass black holes},''
  \href{http://dx.doi.org/10.1016/0550-3213(96)00365-3}{{\em Nucl. Phys.} {\bf
  B476} (1996)  515--547},
\href{http://arxiv.org/abs/hep-th/9604177}{{\tt arXiv:hep-th/9604177
  [hep-th]}}.

\bibitem{Goulart:2016nkv}
P.~Goulart, ``{Massless black holes and charged wormholes in string theory},''
\href{http://arxiv.org/abs/1611.03164}{{\tt arXiv:1611.03164 [hep-th]}}.

\bibitem{Clement:2009ai}
G.~Clement, J.~C. Fabris, and M.~E. Rodrigues, ``{Phantom Black Holes in
  Einstein-Maxwell-Dilaton Theory},''
  \href{http://dx.doi.org/10.1103/PhysRevD.79.064021}{{\em Phys. Rev.} {\bf
  D79} (2009)  064021},
\href{http://arxiv.org/abs/0901.4543}{{\tt arXiv:0901.4543 [hep-th]}}.

\bibitem{Bergmann:1957zza}
O.~Bergmann and R.~Leipnik, ``{Space-Time Structure of a Static Spherically
  Symmetric Scalar Field},''
\href{http://dx.doi.org/10.1103/PhysRev.107.1157}{{\em Phys. Rev.} {\bf 107}
  (1957)  1157--1161}.

\bibitem{Fisher:1948yn}
I.~Z. Fisher, ``{Scalar mesostatic field with regard for gravitational
  effects},'' {\em Zh. Eksp. Teor. Fiz.} {\bf 18} (1948)  636--640,
\href{http://arxiv.org/abs/gr-qc/9911008}{{\tt arXiv:gr-qc/9911008 [gr-qc]}}.

\bibitem{Bronnikov:1973fh}
K.~A. Bronnikov, ``{Scalar-tensor theory and scalar charge},''
{\em Acta Phys. Polon.} {\bf B4} (1973)  251--266.

\bibitem{Ellis:1973yv}
H.~G. Ellis, ``{Ether flow through a drainhole - a particle model in general
  relativity},''
\href{http://dx.doi.org/10.1063/1.1666161}{{\em J. Math. Phys.} {\bf 14} (1973)
   104--118}.

\bibitem{Morris:1988cz}
M.~S. Morris and K.~S. Thorne, ``{Wormholes in space-time and their use for
  interstellar travel: A tool for teaching general relativity},''
\href{http://dx.doi.org/10.1119/1.15620}{{\em Am. J. Phys.} {\bf 56} (1988)
  395--412}.

\bibitem{Hendi:2014uea}
S.~H. Hendi, ``{Wormhole Solutions in the Presence of Nonlinear Maxwell
  Field},'' \href{http://dx.doi.org/10.1155/2014/697863}{{\em Adv. High Energy
  Phys.} {\bf 2014} (2014)  697863},
\href{http://arxiv.org/abs/1405.6997}{{\tt arXiv:1405.6997 [physics.gen-ph]}}.

\bibitem{Nakonieczna:2015apa}
A.~Nakonieczna, M.~Rogatko, and Å.~Nakonieczny, ``{Dark sector impact on
  gravitational collapse of an electrically charged scalar field},''
  \href{http://dx.doi.org/10.1007/JHEP11(2015)012}{{\em JHEP} {\bf 11} (2015)
  012},
\href{http://arxiv.org/abs/1508.02657}{{\tt arXiv:1508.02657 [hep-th]}}.

\bibitem{Gibbons:2008rj}
G.~W. Gibbons and M.~C. Werner, ``{Applications of the Gauss-Bonnet theorem to
  gravitational lensing},''
  \href{http://dx.doi.org/10.1088/0264-9381/25/23/235009}{{\em Class. Quant.
  Grav.} {\bf 25} (2008)  235009},
\href{http://arxiv.org/abs/0807.0854}{{\tt arXiv:0807.0854 [gr-qc]}}.

\bibitem{Bronnikov:1977is} 
  K.~A.~Bronnikov and G.~N.~Shikin,
  ``Interacting Fields in General Relativity,''
  Izv.\ Vuz.\ Fiz.\  {\bf N9 1977}, 25 (1977).
  
\bibitem{Bronnikov:1979ex} 
  K.~A.~Bronnikov, V.~N.~Melnikov, G.~N.~Shikin and K.~P.~Staniukowicz,
  ``Scalar, Electromagnetic, And Gravitational Fields Interaction: Particle - Like Solutions,''
  Annals Phys.\  {\bf 118}, 84 (1979).

\bibitem{Gibbons:1987ps}
G.~W. Gibbons and K.-i. Maeda, ``{Black Holes and Membranes in Higher
  Dimensional Theories with Dilaton Fields},''
\href{http://dx.doi.org/10.1016/0550-3213(88)90006-5}{{\em Nucl. Phys.} {\bf
  B298} (1988)  741--775}.

\bibitem{Garfinkle:1990qj}
D.~Garfinkle, G.~T. Horowitz, and A.~Strominger, ``{Charged black holes in
  string theory},'' \href{http://dx.doi.org/10.1103/PhysRevD.43.3140,
  10.1103/PhysRevD.45.3888}{{\em Phys. Rev.} {\bf D43} (1991)  3140}.
[Erratum: Phys. Rev.D45,3888(1992)].

\bibitem{Jusufi:2017gyu}
K.~Jusufi, ``{Deflection Angle of Light by Wormholes using the Gauss-Bonnet
  Theorem},''
\href{http://arxiv.org/abs/1706.01244}{{\tt arXiv:1706.01244 [gr-qc]}}.

\bibitem{Jusufi:2017vta}
K.~Jusufi, A.~{\"O}vg{\"u}n, and A.~Banerjee, ``{Light Deflection by Charged
  Wormholes in Einstein-Maxwell-Dilaton Theory},''
\href{http://arxiv.org/abs/1707.01416}{{\tt arXiv:1707.01416 [gr-qc]}}.

\bibitem{Jusufi:2015laa}
K.~Jusufi, ``{Gravitational lensing by Reissner-Nordstr\"{o}m black holes with
  topological defects},''
  \href{http://dx.doi.org/10.1007/s10509-015-2609-8}{{\em Astrophys. Space
  Sci.} {\bf 361} (2016) no.~1, 24},
\href{http://arxiv.org/abs/1510.08526}{{\tt arXiv:1510.08526 [gr-qc]}}.

\end{thebibliography}
\end{document}